\documentstyle[sprocl,epsfig]{article}

\def\NPB{{\em Nucl. Phys.} B}
\def\PLB{{\em Phys. Lett.}  B}

\def\PRD{{\em Phys. Rev.} D}
\def\ZPC{{\em Z. Phys.} C}

\def\GeV{\hbox{\rm GeV}}
\def\be{\begin{equation}}
\def\ee{\end{equation}}
\def\bea{\begin{eqnarray}}
\def\eea{\end{eqnarray}}

\begin{document}

\title{LEADING $\ln(1/x)$ AND HEAVY QUARK CORRECTIONS IN STRUCTURE FUNCTIONS} 

\author{R. S. THORNE}

\address{Department of Physics, Theoretical Physics, 1 Keble Road,\\
Oxford OX1 3NP, England\\E-mail: thorne@v2.rl.ac.uk}


\maketitle\abstracts{I present a calculation of structure functions at 
leading order which includes an unambiguous inclusion of the leading 
$\ln(1/x)$ terms for each power of $\alpha_s$, and also the correct effects
due to the mass of the charm and bottom quarks. I compare the results of fits 
to data to those obtained using conventional NLO in $\alpha_s$ 
calculations, noting a clear preference for my approach, especially at 
small $x$. The predictions for both the charm structure function and 
$F_L(x,Q^2)$ using the two approaches are compared, 
the latter being much more discriminating.}

\section{Introduction}

There has recently been a great deal of theoretical activity concerning
the calculation of structure functions at small $x$, driven by the vast 
increase in data in this region obtained at HERA 
\cite{hone,zeus}. One of the main questions is whether one should 
include potentially important leading $\ln(1/x)$ terms at high orders in 
$\alpha_s$, or simply order in powers of $\alpha_s$ alone.
In a number of previous publications I showed that it is possible to include
the leading $\ln(1/x)$ terms within the conventional renormalization 
group approach in an unambiguous manner at leading order \cite{lorsc}. 
The method of 
calculation is based on 3 points. 

\noindent 1. The quantities one calculates to a given order are directly
observable. Hence, I calculate in terms of the structure function at a 
particular scale ($Q_I^2$) and its evolution away from this 
scale.\footnote{This results in expressions in terms of 
Catani's ``physical anomalous dimensions''\cite{catani}.}  

\noindent 2. The leading--order expression for each independent part of a 
physical quantity begins at its lowest power of $\alpha_s$, i.e. if a term 
$\ln^m(1/x)$ first appears at order $\alpha_s^n$ this is the leading order 
for this form of $x$ dependence. 

\noindent 3. The inputs for the structure functions are two flat 
nonperturbative 
functions for $F_L$ and $F_2$, convoluted with calculable perturbative 
contributions. These perturbative parts are determined by demanding that 
the expressions for the structure functions are invariant under changes of 
starting scale order by order in $\alpha_s$, and are hence determined by the 
perturbatively controlled evolution. 

This leads to a 
leading--order--renormalization--scheme--consistent (LORSC) calculation of 
structure functions, which naturally combines the leading $\ln(1/x)$ 
expansion with the more conventional $\alpha_s$ expansion. When a global fit 
was performed the quality using the LORSC
calculation was superior to that using a conventional NLO--in--$\alpha_s$
calculation, particularly at small $x$. 

\section{Implementation of Heavy Quarks}

The main problem with this previous analysis was that it
used a very naive prescription for heavy quarks,
i.e. the charm and bottom quarks were both treated as infinitely massive 
below thresholds $Q^2=m_H^2$, and as massless above these thresholds.
With the direct data on the charm structure function \cite{honec,zeusc,emc},
and the fact that charm comprises $\sim 20\%$ of the total $F_2(x,Q^2)$
at the lowest $x$ values at HERA this is no longer sufficient.  
A method for including the heavy quark contributions in a manner which 
guarantees both smoothness at the  threshold of $W^2\equiv Q^2(x^{-1}-1)=
4m_H^2$ and the correct summation of $\ln(Q^2/m_H^2)$ terms 
was developed for the
conventional approach \cite{rt}. The extension to the LORSC
calculation is in principle quite simple, but in practice rather involved. 
Essentially it involves  imposing matching conditions at $Q^2=m_H^2$ such 
that the evolution is continuous at this point, but in 
terms of effective heavy quark coefficient functions and anomalous 
dimensions\footnote{These involve the leading $\ln(1/x)$
heavy quark coefficient functions \cite{ciaf}.} 
which determine the heavy quark structure function and its 
evolution in terms of the light quark structure functions, rather than parton 
distributions. The details of this LORSC(H) calculation will be presented in 
a future publication.

\begin{figure}
\centerline{\epsfig{figure=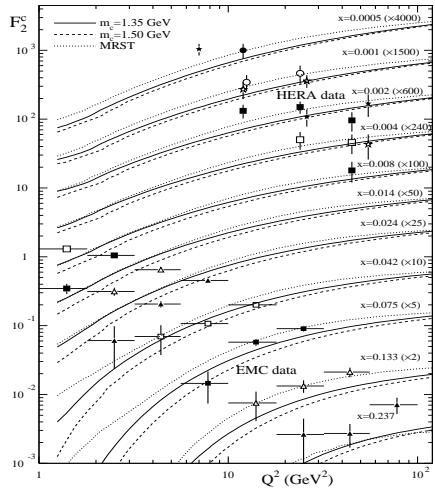,height=3in,width=2.5in}}
\vspace{-0.5in}
\caption{Comparison of $F_{2,c}(x,Q^2)$ with data using the LORSC(H) fit 
with $m_c=1.35\GeV$ and $1.5\GeV$ and the MRST fit ($m_c=1.35 \GeV$).}
\vspace{-0.15in}
\label{fig:charmfig}
\end{figure}

The quality of the LORSC(H) fit to a wide variety of structure function data 
(references can be found in \cite{mrst})
is compared to that for a NLO--in--$\alpha_s$
fit using an analogous treatment of charm, i.e. the recent MRST fit 
\cite{mrst}. The quality of the fit
(using $m_c=1.35\GeV$), in different
$x$ bins is shown in table \ref{tab:chisquare}. The LORSC(H) 
is better overall than the MRST fit, being very slightly worse at high 
$x$, but considerably better at small $x$.  

\begin{table}[t]
\caption{Quality of fit using the LORSC(H) and the 
NLO--in--$\alpha_s$ (MRST) approaches.\label{tab:chisquare}}
\vspace{0.2cm}
\begin{center}
\begin{tabular}{|c|c|c|c|}
\hline
$x$-range & data points & LORSC(H) $\chi^2$ & MRST $\chi^2$\\
\hline
$x\geq$ 0.1 & 597 & 688 & 682\\
0.1$>x\geq$ 0.01 & 385 & 382 & 377\\
$x<$ 0.01 & 278 & 219 & 272\\
\hline
total & 1260 & 1289 & 1332 \\
\hline
\end{tabular}
\end{center}
\vspace{-0.15in}
\end{table}

\begin{figure}
\centerline{\epsfig{figure=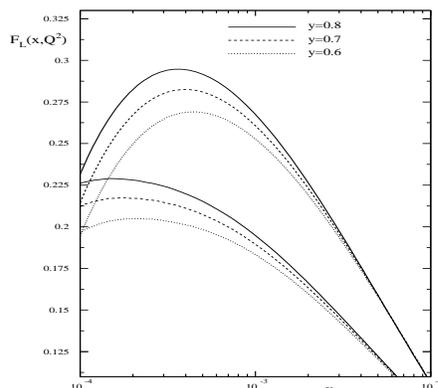,height=2.4in,width=2.4in}}
\vspace{-0.5in}
\caption{Prediction for $F_{L}(x,Q^2)$ in the HERA range 
for the LORSC(H) fit (lower curves) and for the MRST fit (upper curves).}
\vspace{-0.15in}
\label{fig:flfig}
\end{figure}

We can also 
compare the results of the predictions from the two approaches.
Fig. \ref{fig:charmfig} shows the LORSC(H) calculation of 
$F_{2,c}(x,Q^2)$ (using $m_c=1.35 {\hbox{\rm GeV}}$ and $m_c=1.5\GeV$), 
and also the prediction from the MRST fit (where $m_c =1.35\GeV$). Although
in principle the two approaches could give different predictions for 
$F_{2,c}(x,Q^2)$, in practice they 
are rather similar when the same value of $m_c$ is used. 
However, while the MRST fit is very sensitive to $m_c$,
becoming worse quickly when it increases above $1.35\GeV$, the 
LORSC(H) fit is almost unchanged in going from $1.35 \GeV$ to $1.5\GeV$,
and provides much more freedom in $F_{2,c}(x,Q^2)$. 
A larger difference between the approaches is observed when comparing the 
predictions for $F_L(x,Q^2)$, as seen in fig. 
\ref{fig:flfig}. We note that both predictions
for $F_L(x,Q^2)$ are significantly lower than when using
the previous treatment for charm, since in the correct approach the heavy
quark contribution to $F_L(x,Q^2)$ is strongly suppressed until 
$Q^2\gg m_c^2$.

\section{Conclusion}

It appears as though the inclusion of leading $\ln(1/x)$ terms in a 
consistent manner significantly improve the comparison to structure function 
data at small $x$, highlighting the shortcomings of a  
NLO--in--$\alpha_s$ calculation in this region. The calculation of the 
kernel for the 
NLO--in--$\ln(1/x)$ gluon Green's function has recently been completed
\cite{lipfad,ciafcam}, and is very suggestive that the NLO--in--$\ln(1/x)$
corrections are extremely large. Nevertheless, further 
understanding seems necessary before it is known precisely how this 
new result relates to a consistent calculation of structure functions. It may 
indeed  be true that a RSC expansion scheme is not really convergent. 
However, it is 
already known that this is true for the conventional $\alpha_s$ expansion:
the predictions for $F_L(x,Q^2)$ being hugely different at small $x$ at LO
and NLO. Hence, I propose that the sucess of the inclusion of leading 
$\ln(1/x)$ terms in the correct manner is telling us something important
about the true physics at small $x$.

\end{document}